\documentclass{myeas}
\usepackage{graphicx}
\usepackage{eurosym}
\usepackage{hyperref}
\def\farcs{\hbox{$.\!\!^{\prime\prime}$}}

\begin{document}

\title{GYES, a multifibre spectrograph for the CFHT} 
\author{P. Bonifacio}\address{GEPI, Observatoire de Paris, CNRS, Universit\'e Paris Diderot, 5 Place Jules Janssen, 92190, Meudon, France}
\author{S. Mignot}\sameaddress{1}
\author{J.-L. Dournaux}\sameaddress{1}
\author{P. Fran\c{c}ois}\sameaddress{1}
\author{E. Caffau}\sameaddress{1}
\author{F.~Royer}\sameaddress{1}
\author{C. Babusiaux}\sameaddress{1}
\author{F. Arenou}\sameaddress{1}
\author{C. Balkowski}\sameaddress{1}
\author{O. Bienaym\'e}\address{Observatoire astronomique de Strasbourg, 11, rue de l'Universit\'e F-67000 Strasbourg, France} 
\author{D.~Briot}\sameaddress{1}
\author{R. Carlberg}\address{Department of Astronomy and Astrophysics, University of Toronto, Toronto, Ontario M5S 3H4 Canada}
\author{M. Cohen}\sameaddress{1}
\author{G. B. Dalton}\address{University of Oxford,
Astrophysics, Department of Physics,
Keble Road, Oxford,, OX1 3RH, UK}
\author{B. Famaey}\sameaddress{2}
\author{G.~Fasola}\sameaddress{1}
\author{Y. Fr\'emat}\address{Royal Observatory Belgium Astrometry-Dynamics of Celestial Bodies, Av Circulaire 3-Ringlaan 3, BE 1180 Brussels
Belgium} 
\author{A. G\'omez}\sameaddress{1}
\author{I. Guinouard}\sameaddress{1}
\author{M. Haywood}\sameaddress{1}
\author{V.~Hill}\address{Universit\'e de Nice Sophia Antipolis, CNRS,
Observatoire de la C\^ote d'Azur, Laboratoire Cassiop\'e e, B.P. 4229, 06304 Nice Cedex 4, France}
\author{J.-M. Huet}\sameaddress{1}
\author{D. Katz}\sameaddress{1}
\author{D. Horville}\sameaddress{1}
\author{R. Kudritzky}\address{Institute for Astronomy, University of Hawaii at Manoa, 2680 Woodlawn Drive, Honolulu, Hawaii 96822, USA}
\author{R.~Lallement}\address{LATMOS/IPSL, 11 Bd D'Alembert, 78280 Guyancourt, France}
\author{Ph. Laporte}\sameaddress{1}
\author{P. de Laverny}\sameaddress{6}
\author{B. Lemasle}\address{Kapteyn Astronomical Institute, Landleven 12, 9747 AD  Groningen, The Netherlands}
\author{I.J.~Lewis}\sameaddress{4}
\author{C.~Martayan}\address{European Southern Observatory, Alonso de C\`ordova, 3107 Vitacura - Santiago, Chile}\sameaddress{,1}
\author{R. Monier}\address{Universit\'e de Nice Sophia Antipolis, CNRS, Observatoire de la C\^ote d'Azur, Laboratoire Fizeau, B.P. 4229, 06304 Nice Cedex 4. France}
\author{D. Mourard}\sameaddress{11}
\author{N. Nardetto}\sameaddress{11}
\author{A. Recio Blanco}\sameaddress{6}
\author{N.~Robichon}\sameaddress{1}
\author{A.C. Robin}\address{Observatoire de Besan\c con, 41 bis, avenue de l'Observatoire, B.P. 1615, 25010 Besan\c con Cedex, France}
\author{M. Rodrigues}\sameaddress{1}
\author{C. Soubiran}\address{Observatoire de Bordeaux, 2 Avenue de l'Observatoire, F-33270 Floirac, France}
\author{C.~Turon}\sameaddress{1}
\author{K.~Venn}\address{Department of Physics \& Astronomy
The University of Victoria,
Elliott Building, 3800 Finnerty Road
Victoria, BC, V8P 5C2, Canada}
\author{Y. Viala}\sameaddress{1}
\begin{abstract}

We have chosen the name of GYES, one of the mythological giants
with one hundred  arms, offspring of Gaia and Uranus, for our instrument
study of a multifibre spectrograph for the prime focus of
the Canada-France-Hawaii Telescope.
Such an instrument could provide an excellent ground-based
complement for the Gaia mission and a northern complement
to the HERMES project on the AAT.
The CFHT is well known for providing a stable prime focus
environment, with a large field of view,
which has hosted several imaging instruments, 
but has never hosted a multifibre spectrograph. 
Building upon the experience gained at GEPI with FLAMES-Giraffe
and X-Shooter,
we are investigating the feasibility of a high multiplex
spectrograph (about 500 fibres) over a field of view  1 degree in
diameter.
We are investigating an instrument with resolution in the 
range 15\,000 to 30\,000, which should provide accurate chemical
abundances for stars down to 16th magnitude and radial 
velocities, accurate to 1 $\rm kms^{-1}$ for fainter stars.
The study is led by GEPI-Observatoire de Paris with a contribution
from Oxford for the study of the positioner. The financing for the
study comes from INSU CSAA and Observatoire de Paris. 
The conceptual study will be delivered to CFHT for review by October 1st 2010.
\end{abstract}
\maketitle
\section{Introduction}

The HIPPARCOS mission was highly successful in providing
parallaxes and proper motions, but right from the beginning
of the project it was realised
that the lack of matching radial velocities seriously
hampered the study of Galactic structure and dynamics.
Unfortunately the proposals to build dedicated ground based
telescopes for the radial velocity measurements were not 
financed.
Long term programmes on existing facilities have strived
to obtain the radial velocities (Fehrenbach et al.
\cite{Fehrenbach} and references therein, Grenier et al. 
\cite{Grena,Grenb}, Nordstr\"om et al. \cite{GCS}),
however a compilation of these and other sources, only
amounts to radial  velocities for 35\,495 stars (Gontcharov
\cite{Gont06}), which represents  30\% of the Hipparcos catalogue.

Building upon this experience, from the beginning of the study of
Gaia  it was planned that it should include an instrument
capable of measuring radial velocities. 
Since to measure a radial velocity one needs a spectrum, there is
a lot more information than radial velocity that can be 
be extracted from a spectrum: effective temperatures, surface
gravities, chemical abundances and rotational velocities.
The clever phase A design  of the Radial Velocity Spectrometer (RVS,
Katz et al. \cite{Katz}) allowed to obtain this and provide
spectroscopic information essentially for all stars down to magnitude
17.5 (Wilkinson et al. \cite{Wil}). 
As the design of Gaia became more detailed in phases B and C, the
performances of the RVS degraded significantly as a result
of the result of the trade-offs made by the industrial
consortium Astrium and ESA.  

Under these conditions it is of high interest 
to perform ground-based observations which may complement the informations
provided by Gaia, at least for a limited sample of stars.
With this in mind we started  a study for a high resolution
multi-fibre spectrograph, to be  placed at the prime focus
of the Canada-France-Hawaii telescope. In this respect, GYES
is truly a son of Gaia.

\section{Science Requirements}

GYES aims at providing radial 
velocities and chemical abundances for all of the Galactic populations
(thin, thick disks, bulge, halo, open and globular clusters).
The main emphasis is on the  ``chemical labelling'', whereby
we wish to measure chemical abundances with an accuracy of
the order of 0.1\,dex, looking for the characteristic patterns
which allow to separate the various populations.
The radial velocities accuracy ought to  match that provided 
by Gaia for the the transverse velocities,  about 1\,$\rm kms^{-1}$.
We are preparing a 
Survey Strategy document, which concretely illustrates how
GYES can be used to conduct the desired survey.
We target a survey of 50 nights/year over 5 years to cover a
total of more than 800 fields. Main targets 
will be F--G dwarfs and subgiants for which we will be able
to derive ages using the distances provided by Gaia, 
and giant stars to study the gradients at larger distances.
The need for accurate
abundances for several elements, drives the 
request on resolution which should not be lower than  15\,000
and, if possible, as high as 30\,000.

\section{Overview of the baseline}

A baseline has been established starting
from the Science Requirements and making a number
of trade-offs to ensure: 
{\em i)}   maximum efficiency;
{\em ii)}  hardware cost within an envelope of 5M\EUR ;
{\em iii)} simplicity and robustness.
For the field corrector the baseline is to use
the triplet, built by REOSC, that was previously used for direct 
imaging at the prime focus with photographic plates and also
with the electronic camera. 
The unvignetted field-of-view has a diameter of 0.86$^\circ$,
it is however possible to use  a field-of-view of 1.00$^\circ$,
in which the vignetting leads to a loss of only 5\% of light on
the outer ring. 

The fibre positioner unit resembles that of 2dF, consisting of two
plates (with associated sets of fibres) which can be exchanged by tumbling,
orthogonally to the  telescope's axis. The positioning is performed
on the top plate, while the bottom plate is observing, by a commercial
off-the-shelf $x-y$ robot.

The field-corrector and positioner are supported by 
a dedicated Upper End which, when not in use, can be parked
on the floor of the dome, as is currently done with the three 
Upper Ends in use at CFHT (secondary mirror for cassegrain instruments,
MegaPrime-MegaCam, and WIRCam). 
Experiments with the positioning software of 2dF allowed to conclude
that it is possible to position 500 fibres in the available 
one degree field.
The fibres are collected into a single 
fibre bundle that links the focal plane to the slit of spectrograph. 
Two  identical slits are foreseen, one for each observing plate,
like for Giraffe, and a slit-exchange mechanism.
The fibre aperture on the sky will be a micro-lens of
1\farcs{2} diameter. This diameter allows to collect over 90\% of
the light for a typical seeing of 0\farcs{8}, taking into
account also image quality provided by the field corrector. 

The spectrograph itself is placed in a thermally controlled room 
on the floor of the dome.  An actual experiment performed at CFHT, 
of the manoeuvre necessary to unmount
the GYES Upper End, with a rope instead of the fibre bundle, allowed
to conclude that the manoeuvre is itself feasible, without 
breaking the fibre link and that the length
of the fibres, including the necessary slack, is about 50m.

At the exit of the slit the beam is collimated and 
split in two by a dichroic which delivers a blue  and a red 
beam. The two arms of the spectrograph (blue and red)
have a very similar optical design. The dispersing element
is a pair of Volume Phase Holographic (VPH) gratings, used at moderate
angles (around 36$^\circ$), in a configuration
similar to that suggested for the first time
by Span\`o and Bonifacio (\cite{SB}, but see also 
Epps et al. \cite{Epps}).
The advantage of this solution, over that of using a single
VPH grating at higher incidence angles, is that one can obtain 
a very high efficiency together with a very large spectral range.
The dispersed beam is focused by a fully dioptric camera,
at the focal plane of which is a mosaic of two 4k$\times$4k CCDs,
the baseline is E2V CCD231-84 detectors, available off-the-shelf.
This configuration allows to cover the ranges
390-450\,nm in the blue arm and 587-673\,nm in the red.
Each arm has
a  wavelength gap smaller than 3\,nm due to the gap between the CCDs. 
One disadvantage of this optical design is that the resolution 
varies continuously across the range, it is roughly 16\,000 at the
blue end of each range and 30\,000 at the red end.
This is undoubtedly a complication in the interpretation of the data,
however simulations indicate that, as long as the resolution
as a function of wavelength is well known, this is not a major problem.

\section{Conclusions}

At the time of writing the estimates of costs and efficiency for GYES
are not yet consolidated. It nevertheless appears likely 
that the instrument can be built within an envelope of hardware
cost of 5 M\EUR .  The progress of the project and
updated information can be found on its web site
\url{http://gyes.obspm.fr}.


\end{document}